\documentclass[pss]{wiley2sp} 
\usepackage{amsmath}
\usepackage{amssymb}

\begin{document}

\title{Effect of chemical substitution and pressure on YbRh$_2$Si$_2$}

\titlerunning{Effect of chemical substitution and pressure on YbRh$_2$Si$_2$}


\author{M. Nicklas\textsuperscript{\Ast},
 M. E. Macovei,
 J. Ferstl,
 C. Krellner,
 C. Geibel,
 F. Steglich}

\authorrunning{M. Nicklas et al.}

\mail{e-mail
  \textsf{nicklas@cpfs.mpg.de}, Phone:
  +49-351-4646 2400, Fax: +49-351-4646 2402}

\institute{%
 Max Planck Institute for Chemical Physics of Solids,
N\"{o}thnitzer, Strasse 40, 01187 Dresden, Germany}

\received{XXXX, revised XXXX, accepted XXXX} 
\published{XXXX} 


\abstract{%
%
%
%
\abstcol
{%
  We carried out electrical resistivity experiments on (Yb,La)Rh$_2$Si$_2$ and on
    Yb(Rh,Ir)$_2$Si$_2$ under pressure and
    in magnetic fields. YbRh$_2$Si$_2$ exhibits a weak antiferromagnetic transition at atmospheric
    pressure with a N\'{e}el temperature of only $T_N\approx 70$~mK.
    By applying a small magnetic field $T_N$ can be
    continuously suppressed to $T=0$ at $B_c = 60$~mT ($B\bot c$) driving the system
    to a quantum critical point (QCP).
    On applying external pressure the magnetic phase is stabilized and $T_N(p)$ is increasing as
    usually observed in Yb-based heavy-fermion metals.
    }
    {
    Substituting Yb by La or Rh by Ir allows to create a negative
    chemical pressure, La (Ir) being smaller than Yb (Rh), and eventually
    to drive YbRh$_2$Si$_2$ to a pressure controlled QCP.
    In this paper we compare the
    effect of external hydrostatic pressure and chemical substitution
    on the ground-state properties of YbRh$_2$Si$_2$.
  }
}
%
%

%

\maketitle   

\section{Introduction}

New unconventional states are emerging in the vicinity of quantum critical points (QCP), where
critical quantum fluctuations determine the low-temperature properties. Heavy-fermion (HF) metals are
ideal model systems to study quantum-critical behavior. External pressure, chemical substitution, or
magnetic field allow to tune the ground state properties. If a magnetically ordered state is
suppressed continuously to zero temperature, eventually a QCP is reached. In Ce-based HF metals
application of external pressure suppresses magnetism, while in Yb-based materials magnetism is
stabilized.

YbRh$_2$Si$_2$ orders antiferromagnetically at a N\'{e}el temperature of only $T_N\approx70$~mK
\cite{Trovarelli00}. $T_N$ can be suppressed to zero temperature by applying a small magnetic fields
at a critical field $B_{c0}$ of about 0.06 and 0.66~T for $B\perp c$ and $B\parallel c$,
respectively. In the vicinity $B_{c0}$ pronounced non-Fermi liquid behavior is observed
\cite{Gegenwart02}. The QCP in most HF systems is of 3 dimensional spin-density wave (conventional)
type. At a conventional QCP the quasiparticles formed by the 4$f$ electrons and the conduction
electrons stay intact. Recently, the discovery of an unconventional QCP in YbRh$_2$Si$_2$ attracted
special attention \cite{Custers03,Paschen05}. In YbRh$_2$Si$_2$ it is discussed that the
quasiparticles decompose at the QCP. This is related to a break down of the Kondo effect, introducing
an additional energy scale, $T^*(B)$. The break down of the Kondo effect is reflected in a change of
the Fermi volume, which has been inferred from results of the Hall coefficient, ac-susceptibility,
specific heat, magnetostriction, and magnetoresistance \cite{Paschen05,Gegenwart07}. Here, we present
a substitution and pressure study of YbRh$_2$Si$_2$.

\section{Experimental Details}

Single crystals of Yb$_{1-x}$La$_x$\\Rh$_2$Si$_2$ ($x=0.05$ and $0.1$) and of
Yb(Rh$_{1-y}$Ir$_y$)$_2$Si$_2$ ($y=0.06$) were grown in In-flux. The ThCr$_2$Si$_2$ structure was
confirmed by X--ray powder diffraction using ground single crystalline material. The stoichiometry of
the samples was verified using energy dispersive X-ray analysis (EDX). Electrical resistivity
($\rho$) measurements under hydrostatic pressure ($p$) have been carried out by a standard four-probe
technique. In different piston-cylinder type cells pressure up to 2.5~GPa was generated. The shift of
the superconducting transition temperature of lead (tin) under pressure served as pressure gauge. In
a $^3$He--$^4$He dilution cryostat equipped with a superconducting magnet temperatures down to 50~mK
and magnetic fields up to 8~T could be reached.

\section{Results and Discussion}

The temperature--pres\-sure ($T-p$) phase diagram of Yb$_{1-x}$La$_x$Rh$_2$Si$_2$  and \linebreak
Yb(Rh$_{1-y}$Ir$_y$)$_2$Si$_2$ was determined by electrical resistivity experiments. At ambient
pressure none of the investigated samples, Yb$_{1-x}$La$_x$Rh$_2$Si$_2$ ($x=0.05$ and $0.1$) and
Yb(Rh$_{1-y}$Ir$_y$)$_2$Si$_2$ ($y=0.06$), shows the onset of magnetic order down to $T=50$~mK. Both
La- and Ir-substitution are expanding the crystal lattice leading to a negative chemical pressure
and, therefore, are expected to suppress the magnetism observed in YbRh$_2$Si$_2$. Replacing 5\%
(10\%) Yb by La expands the unit cell volume by about $\sim0.2$\% ($\sim0.4$\%). Using the bulk
modulus of YbRh$_2$Si$_2$, $B=187$~GPa \cite{Plessel03}, as an approximation for the substituted
compounds, the hypothetical (negative) pressure corresponding to the lattice expansion can be
calculated as $\Delta p= - (B/V)\Delta V$, where $V$ is the unit-cell volume. For $x=0.05$ ($x=0.1$)
a corresponding pressure $\Delta p=-0.35$~GPa ($\Delta p=-0.67$~GPa) is obtained. In the same way in
Yb(Rh$_{1-y}$Ir$_y$)$_2$Si$_2$ the substitution of Rh by Ir, $y=0.06$, expands the lattice by
$\sim0.03$\% which corresponds to $\Delta p=-0.06$~GPa.

The resistivity of the La- and Ir-substituted YbRh$_2$Si$_2$ exhibits the typical temperature
dependence of a Kondo-lattice system. At ambient pressure $\rho(T)$ shows a clear maximum around
$T_M\approx143$~K for both La substituted compounds \cite{Ferstl05}, while $T_M\approx130$~K for the
6\% Ir substituted sample \cite{Macovei08}. Below $T_M$ $\rho(T)$ decreases strongly due to the onset
of coherent Kondo scattering. With increasing pressure $T_M(p)$ shifts to lower temperatures with an
initial rate of about ${\rm d}T_M/{\rm d}p\,|_{p=0}\approx-31$~K/GPa for the La-substituted samples
and ${\rm d}T_M/{\rm d}p\,|_{p=0}\approx-46$~K/GPa for Yb(Rh$_{0.94}$Ir$_{0.06}$)$_2$Si$_2$. These
values are comparable with the one found in YbRh$_2$Si$_2$ \cite{Mederle02}. $T_M(p)$ can be taken as
a measure of the pressure dependence of the Kondo temperature. Therefore, a decreasing $T_M$
evidences a weakening of the Kondo-interactions. In other words, the system is moved toward magnetism
by application of pressure, as it is expected in a simple Doniach picture for Yb-based HF metals.

\begin{figure}[t]%
\center
\includegraphics*[width=1\linewidth]{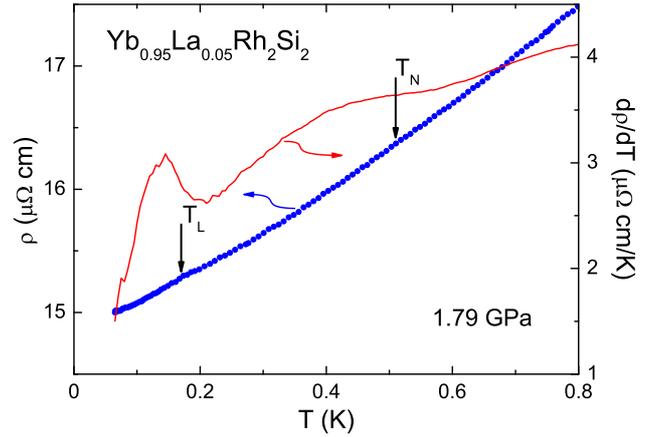}\hfill
\caption{%
$\rho(T)$ (left axis) and ${\rm d}\rho(T)/{\rm d}T$ (right axis) of
Yb$_{0.95}$La$_{0.05}$Rh$_2$Si$_2$ at $p=1.79$~GPa. The arrows indicate $T_N$ and $T_L$,
respectively.} \label{YLRS_rho(T)}
\end{figure}

\begin{figure}[b]%
\center
\includegraphics*[width=1\linewidth]{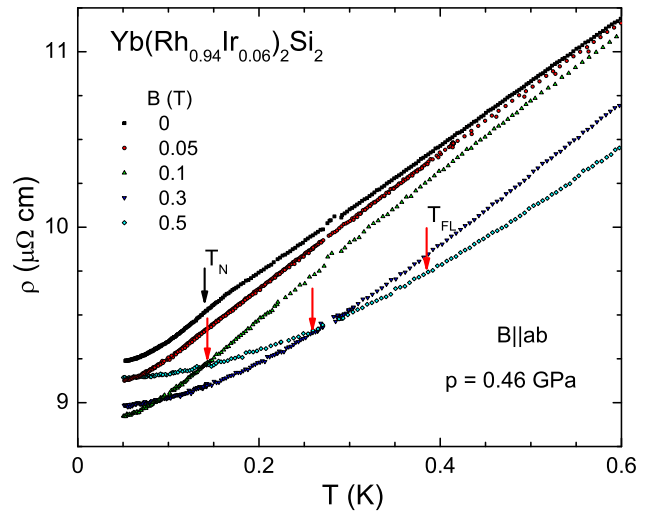}\hfill
\caption{%
$\rho(T)$ of Yb(Rh$_{0.94}$Ir$_{0.06}$)$_2$Si$_2$ at $p=0.46$~GPa in different magnetic fields. The
 arrow indicates the
 N\'{e}el transition at $B=0$. At $B=0.1$, 0.3 and 0.5~T the Fermi-liquid crossover temperature $T_{FL}$
 is marked by arrows.}
\label{YRIS_rho(T)}
\end{figure}

\begin{figure}[t]%
\center
\includegraphics*[width=1\linewidth]{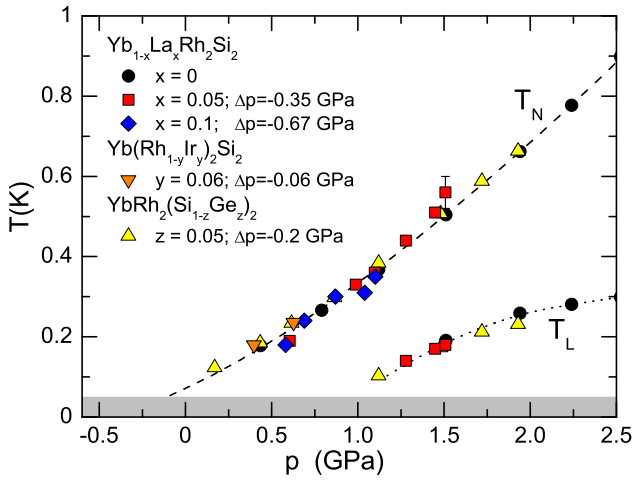}
\caption{%
  $p-T$ phase diagram of YbRh$_2$Si$_2$ summarizing the pressure studies on the La, Ir, and Ge substituted samples.
  The data of the substituted samples are shifted on the pressure axis with respect to pure YbRh$_2$Si$_2$
  according pressure shift given in the figure. For Yb$_{0.95}$La$_{0.05}$Rh$_2$Si$_2$ a representative error bar
  is given for $T_N$; for $T_L$ the symbol size is larger than the error.
  Data on YbRh$_2$Si$_2$ and YbRh$_2$(Si$_{0.95}$Ge$_{0.05}$)$_2$ were taken from Ref.~\cite{Mederle02} and the data on Yb$_{0.9}$La$_{0.1}$Rh$_2$Si$_2$
  from Ref.~\cite{Nicklas06}. See text for details.}
\label{p-T_phase_diagram}
\end{figure}

At ambient pressure $\rho(T)$ of the investigated \linebreak Yb$_{1-x}$La$_x$Rh$_2$Si$_2$ samples
shows a quasi-linear temperature dependence \cite{Weickert06}. This behavior is also observed in the
low pressure range. In YbRh$_2$Si$_2$ a clear drop of the resistance is observed at $T_N\approx70$~mK
\cite{Trovarelli00}. For neither of the two different Yb$_{1-x}$La$_x$Rh$_2$Si$_2$ concentrations a
similar drop is observable in the pressure range, where it would be expected according to our
previous considerations. In a more detailed analysis, however, ${\rm d}\rho/{\rm d}T$ shows a broad
feature indicating $T_N$. $T_N$ is taken as the temperature where the derivative changes its
curvature. In Yb$_{0.95}$La$_{0.05}$Rh$_2$Si$_2$ it is first observed at $p=0.95$~GPa at
$T_N=0.19$~K. With further increasing pressure $T_N(p)$ shifts to higher temperatures. At 1.62 GPa a
second feature at $T_L<T_N$ is observed in ${\rm d}\rho/{\rm d}T$ which is also visible as a small
drop in $\rho(T)$. Figure~\ref{YLRS_rho(T)} displays $\rho(T)$ and its derivative at $p=1.79$~GPa.
The kink at $T_L$ corresponds to a second magnetic transition below $T_N$ which has been also
reported in YbRh$_2$Si$_2$ under pressure \cite{Mederle02}. The transition at $T_L$ has been assigned
to a reorientation of the magnetic moments. The reason for the very weak feature of the magnetic
transition in the resistivity data of La-substituted YbRh$_2$Si$_2$ might be related to the
relatively high value of the residual resistivity, $\rho_0$, compared with YbRh$_2$Si$_2$. However,
in Yb(Rh$_{0.94}$Ir$_{0.06}$)$_2$Si$_2$ at $p=0.46$~GPa the transition at $T_N$ is clearly visible
(see Fig.~\ref{YRIS_rho(T)}) with a residual resistivity comparable with that of the 10\% La
substituted sample. Furthermore, at higher pressures both magnetic transitions at $T_N$ and $T_L$ are
visible as two successive kinks in $\rho(T)$ \cite{Macovei08}.

The $p-T$ phase diagram presented in Fig.~\ref{p-T_phase_diagram} summarizes the pressure studies on
Yb$_{1-x}$La$_x$Rh$_2$Si$_2$ and Yb(Rh$_{1-y}$Ir$_y$)$_2$Si$_2$. Additionally, data obtained on
\linebreak YbRh$_2$Si$_2$ and YbRh$_2$(Si$_{1-z}$Ge$_z$)$_2$ are included \cite{Mederle02}. $T_N(p)$
and $T_L(p)$ for the corresponding samples are shifted according to the effective chemical pressure,
which has been calculated as explained before. As a result $T_N(p)$ and $T_L(p)$, respectively, fall
each on one single line evidencing the existence of a generalized $p-T$ phase diagram. This
highlights a close correspondence between chemical pressure and external pressure.

\begin{figure}[t]%
\center
\includegraphics*[width=1\linewidth]{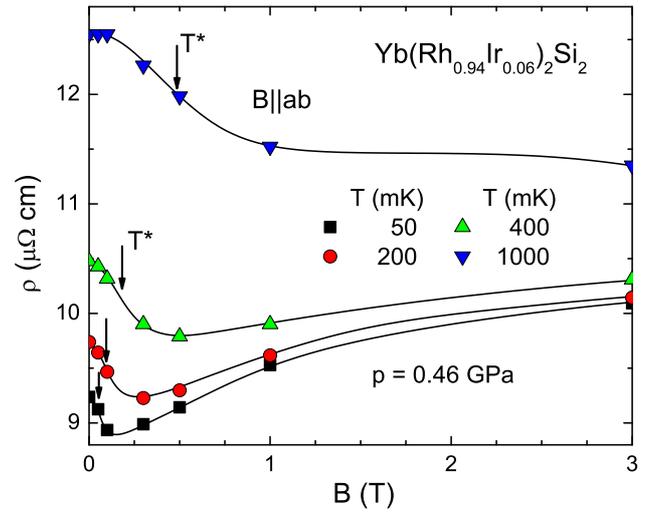}%
\caption{%
 $\rho(H)$ of Yb(Rh$_{0.94}$Ir$_{0.06}$)$_2$Si$_2$ at an external pressure $p=0.46$~GPa
 for selected temperatures deduced from the data presented in Fig.~\ref{YRIS_rho(T)}. The lines are guides to the eye.
 The arrows indicate $T^*$.}
\label{YRIS_rho(H)}%
\end{figure}

\begin{figure}[h]%
\center
\includegraphics*[width=1\linewidth]{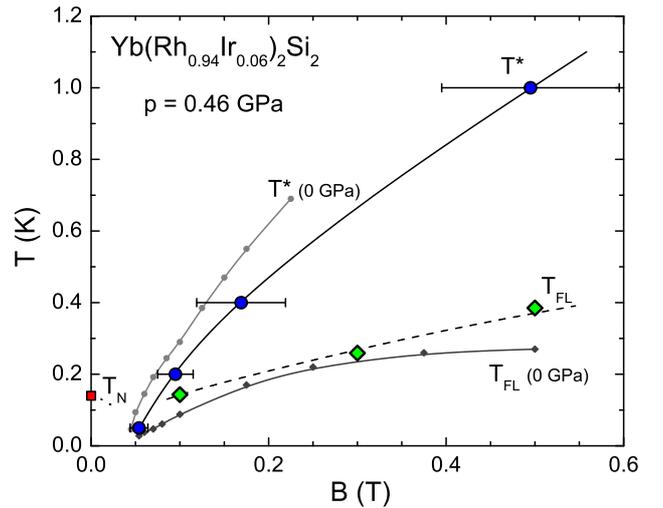}
\caption{%
 $B-T$ phase diagram of  Yb(Rh$_{0.94}$Ir$_{0.06}$)$_2$Si$_2$ at $p=0.46$~GPa.
 Additionally, data at $p=0$ is included \cite{Friedemann09}.} \label{YRIS_phd}
\end{figure}

In the following, we will discuss the $B-T$ phase diagram of Yb(Rh$_{0.94}$Ir$_{0.06}$)$_2$Si$_2$ and
its pressure evolution. At atmospheric pressure no magnetic order is found down to 20~mK
\cite{Westerkamp08}. An extrapolation from our pressure experiments gives a value of about
$T_{N}\approx 20$ mK \cite{Macovei08}, consistent with the results of the experiments at ambient
pressure. A pressure of $p=0.46$~GPa moves the $T_{N}$ up to $0.14$~K, which is twice as large as
$T_{N}$ of YbRh$_2$Si$_2$. Figure \ref{YRIS_rho(T)} displays the low temperature resistivity of
Yb(Rh$_{0.94}$Ir$_{0.06}$)$_2$Si$_2$ at $p=0.46$~GPa for different magnetic fields ($B\perp c$). At
$B=0$ a clear drop in $\rho(T)$ indicates $T_{N}$. In a field of $B=0.05$~T no feature in $\rho(T)$
is visible anymore. From a simple comparison with YbRh$_2$Si$_2$ one would expect a critical field
$B_c\approx100$~mT. By contrast, our results suggest that $T_{N}$ is already suppressed to $T<50$~mK
at 50~mT. In magnetic fields $B\geq0.1$~T, $\Delta\rho\propto T^2$ below the crossover temperature
$T_{FL}$ indicates Landau Fermi liquid behavior. Figure \ref{YRIS_rho(H)} shows $\rho$ as a function
of the magnetic field. The $\rho(B)$ data was extracted from the temperature dependent measurements
displayed in Fig.~\ref{YRIS_rho(T)}. At all investigated temperatures ($T\leq1$~K) $\rho(B)$
initially decreases showing a step-like behavior. At ambient pressure in YbRh$_2$Si$_2$
\cite{Gegenwart07}, as well as in Yb(Rh$_{0.94}$Ir$_{0.06}$)$_2$Si$_2$ \cite{Friedemann09}, the
inflection points of the magnetoresistance measured at different temperatures fall on the same
$T^*(B)$ line determined from the magnetostriction, magnetization, ac-susceptibility, and Hall effect
and ascribed to the Kondo-breakdown crossover. Thus, we take the inflection point of $\rho(B)$ to
determine $T^*(B)$. The resulting $B-T$ phase diagram is shown in Fig.~\ref{YRIS_phd}. Within the
error bars the $T^*(B)$ line at atmospheric pressure \cite{Friedemann09} and at $p=0.46$ are in
reasonable agreement. This holds true also for the Fermi-liquid crossover $T_{FL}(B)$. Our results
suggest the existence of a Kondo breakdown also in Yb(Rh$_{0.94}$Ir$_{0.06}$)$_2$Si$_2$ under
pressure. Further investigations have to be carried out to clarify the details of the phase diagram.

\section{Summary}

We could show that substitution of YbRh$_2$Si$_2$ by small amounts of La, Ir, and Ge in general acts
as negative chemical pressure. We could further establish a generalized phase diagram with the
unit-cell volume as the only free parameter. In the $B-T$ phase diagram of
Yb(Rh$_{0.94}$Ir$_{0.06}$)$_2$Si$_2$ at $p=0.46$~GPa we could establish the existence of the $T^*(B)$
line.




%
%

\end{document}